\documentclass[aps,preprint,prd,epsfig,axodraw,nofootinbib]{revtex4-1}
\usepackage{amssymb}
\usepackage{amsmath}
\usepackage{graphicx}
\usepackage{fancyhdr}
\usepackage{epsfig}
\def\br{\begin{eqnarray}}
\def\er{\end{eqnarray}}
\def\be{\begin{equation}}
\def\ee{\end{equation}}

\def\({\left(}
\def\){\right)}

\def\lesssim{\mathrel{\hbox{\rlap{\hbox{\lower4pt\hbox{$\sim$}}}\hbox{$<$}}}}
\def\gtrsim{\mathrel{\hbox{\rlap{\hbox{\lower4pt\hbox{$\sim$}}}\hbox{$>$}}}}
\begin{document}

\title{Electric charge quantization in $331$ Models with exotic charges}


\author{David Romero Abad\footnote{davidromeroabad@gmail.com},
Jose Reyes Portales\footnote{josereyes2201@gmail.com}, Elmer Ramirez Barreto\footnote{elmerraba@gmail.com} }
	
\affiliation{{ Grupo de Investigaci\'on en F\'isica, Universidad San Ignacio de Loyola\\ Av. La Fontana 550,
	La Molina, Lima, Per\'u.}}







\begin{abstract}
The extensions of the Standard Model based on the $SU(3)_{C} \otimes SU(3)_{L} \otimes U(1)_{X}$ gauge group are known as 331 Models. 
Different properties such as the fermion assignment and the electric charges of the exotic spectrum, that defines a particular 331 model,
are fixed by a $\beta$ parameter. In this article we study the electric charge quantization in two versions of the 331 models,
set by the conditions  $\beta=1/\left( 3\sqrt{3}\right)$ and $\beta=0$. In these frameworks, arise exotic particles, for instance,
new leptons and gauge bosons with a fractional electric charge. Additionally, depending on the version, quarks with non-standard 
fractional electric charges or even neutral appear. Considering the definition of electric charge operator as a linear combination of 
the group generators that annihilates the vacuum, classical constraints from the invariance of the lagrangian, and gauge
and mixed gauge-gravitational anomalies cancellation, the quantization of the electric charge can be verified in both versions. 
\end{abstract}







\maketitle

\section{Introduction}
The experimental observation where the electric charge of the particles appears only in quantized units is known as the electric charge quantization.
The first efforts to explain this phenomenon included ideas related to higher dimensions \cite{Klein},\cite{Arun}, magnetic monopoles \cite{Dirac},
and grand unified theories \cite{Pati}. However, these approaches have not yet been experimentally verified. In this work, we follow the perspective introduced by \cite{Mohapatra}, \cite{Foot1}, \cite{Pisano} and \cite{Pereyra}. This approach considers two main 
conditions, first, classical constraints imposed by the $U(1)_{X}$ gauge group invariance of the Yukawa lagrangian and second, the quantum
restrictions arising from anomalies cancellation. 
In that sense, the above approach for electric charge quantization will be applied for two versions of the $331$ model, set by the conditions $\beta=1/\left( 3\sqrt{3}\right)$ \cite{elmo} and $\beta=0$ \cite{simplest}. In these versions,  new leptons with charge $\pm 2/3\,e$ ($\pm 1/2\,e$), extra quarks with charges $+1/3\,e$, $0$ ($\pm 1/6\,e$),  exotic gauge and scalar bosons with charges $\pm 1/3\,e$ and $\pm 2/3\,e$ ($\pm 1/2\,e$) arise, in addition to a new neutral boson $Z^\prime$.

Also, particles with fractional electric charges have been proposed by other theoretical models \cite{Langa1,Davidson1} and both ATLAS and CMS collaborations have already performed searches of new heavy lepton-like particles with non-standard electric charges \cite{ATLASmono,ATLASlong}. Experimentally, these kinds of particles may be misidentified or unobserved since charged particle identification algorithms generally assume that particles have charges of $\pm 1e$ \cite{Golling}.
The new proposal maintains the special features of the $331$ models, such as the relation between the number of fermionic families and the number of colors in QCD. Reference \cite{Pereyra} considers the quantization of the electric charge in a similar fashion but for the minimal $331$ model, that is, for $\beta=-\sqrt{3}$. In this work, the authors argue that for models with $SU(3)_{C} \otimes SU(3)_{L} \otimes U(1)_{X}$ symmetry, the quantization of the electric charge is verified when they take into account the three families of fermions, considering or not the neutrino masses, in contrast to the SM.  On the other hand, reference \cite{Long} explains the charge quantization using the general form of the electromagnetic currents under the parity invariance for $331$ models with $\beta=-\sqrt{3}$ and $\beta=-1/\sqrt{3}$. Finally, the objective of this study is to verify if the proposed 331 versions, satisfy the quantization of the electric charge.

\section{The Models}
The general relation for the electric charge operator ($\mathcal{Q}$) in a $331$ model is given by
\begin{equation}\label{charge}
\mathcal{Q} = \alpha \  T_3 + \beta \ T_8 + \gamma \  X 
\end{equation}
where $T_3$ and $T_8$ are the diagonal generators of ${\rm SU}(3)_L$ built  as $T_i=\frac{\lambda_i}{2}$ from the Gell-Mann matrices $\lambda_i$, with $i =1,\,...,\,8$, and $X$ is the charge of $U(1)_{X}$.
We consider $\alpha=1$ in order to properly set the  $W$ boson electric charge in the model, as was described in \cite{Long}. \\
In addition,  the value of the $\beta$ parameter fixes the fermion assignment, the electric charges of the exotic spectrum and it is used to classify different 331 models \cite{Martinez}.
We can write:
\begin{eqnarray}\label{charge}
\mathcal{Q}&=&\text{diag} \left( \pm \dfrac{1}{2}\left( 1 + \dfrac{\beta }{\sqrt{3}}  \right), \dfrac{1}{2}\left(\mp 1 \pm \dfrac{\beta }{\sqrt{3}}  \right) , \mp \dfrac{1}{3}\:\beta  \sqrt{3}\right)\nonumber  \\&+&\gamma X I_{3\times 3}
\end{eqnarray}
where the upper sign corresponds to the fundamental representation of the Gell-Mann matrices and the lower sign to the conjugate representation.
The Scalar sector,
\begin{equation}\label{Higgs}
\eta \sim   \left( {\bf 1},{\bf 3}, X_{\eta} \right), \quad 
\rho \sim   \left( {\bf 1},{\bf 3}, X_{\rho}\right), 
\quad 
\chi \sim   \left( {\bf 1},{\bf 3}, X_{\chi} \right)
\end{equation}
which develop vacuum expectation value (vev) as:
\begin{eqnarray}\label{Higgs2}
\left\langle \eta^{0} \right\rangle &=& \dfrac{1}{\sqrt{2}}\left(
\begin{array}{c}
v_{\eta} \\
0  \\
0
\end{array}
\right), \quad 
\left\langle \rho^{0}\right\rangle= \dfrac{1}{\sqrt{2}} \left(
\begin{array}{c}
0 \\
v_{\rho} \\
0
\end{array}
\right), \nonumber \\ 
\quad 
\left\langle \chi^{0}\right\rangle &=& \dfrac{1}{\sqrt{2}} \left(
\begin{array}{c}
0 \\
0  \\
v_{}\chi
\end{array}
\right)
\end{eqnarray}
With the requirement that the charge operator must annihilate the vacuum, we obtain
the following general relations:
{\small 
	\begin{eqnarray}
	\gamma&=& -\dfrac{1}{2}\left( 1+\dfrac{\beta}{\sqrt{3}}\right) \dfrac{1}{X_{\eta}} =-\dfrac{1}{2}\left( -1+\dfrac{\beta}{\sqrt{3}}\right) \dfrac{1}{X_{\rho}}=\dfrac{\beta \sqrt{3}}{3} \dfrac{1}{X_{\chi}} \nonumber \\
	\end{eqnarray}}
Thus, for  the particular choice of  $\beta=1/\left(3\sqrt{3}\right)$, we obtain:
\begin{eqnarray}\label{Higgsvalues1}
\gamma= \dfrac{1}{9 X_{\chi}},\ X_{\eta}=-5X_{\chi}, \ X_{\rho}=4X_{\chi}
\end{eqnarray}
and for  $\beta=0$:
\begin{eqnarray}\label{Higgsvalues2}
\gamma= -\dfrac{1}{2 X_{\eta}}, \ X_{\rho}=-X_{\eta}, \ X_{\chi}=0.
\end{eqnarray}
It is straightforward to verify that the scalar fields fulfill the relation
\begin{eqnarray}\label{Higgsvalues}
X_{\rho} + X_{\eta} +  X_{\chi}=0
\end{eqnarray}
In order to cancel anomalies associated with $SU(3)_{L}$ gauge group, the leptons and the quark families must be  assigned in different
$SU(3)_{L}$ representations. So, for $\beta=1/\left(3\sqrt{3}\right)$, the leptons and the third quark family are assigned in triplets,
while the first two quark families in anti-triplets.\\
The leptonic sector includes:
\begin{eqnarray}
\psi_{i L} &=&  \left(\nu_i,e^{-}_i,E_i\right)_{L}^T
\ \sim \left( {\bf 1},{\bf 3}, X_{\ell_{i}} \right),\nonumber \\  
e_{i_{R}}^{-} &\sim&   \left( {\bf 1},{\bf 1}, X_{e_{i}} \right)   ,\quad  E_{i_{R}} \sim   \left( {\bf 1},{\bf 1}, X_{E_{i}} \right)
\end{eqnarray}
where $i=1,2,3$. \\
The two first quark families form $SU(3)_L$ anti-triplets
\begin{eqnarray}
Q_{a L} &=&  \left(d_a,-u_a,D_{a}\right)_{L}^T
\ \sim \left({\bf 3}, {\bf 3^*}, X_{Q_{a}} \right), \nonumber \\
u_{a_{R}} &\sim& \left({\bf 3}, {\bf 1}, X_{u_{a}} \right), \quad  
d_{a_{R}} \sim \left({\bf 3}, {\bf 1}, X_{d_{a}} \right),\nonumber \\
D_{a_{R}} &\sim& \left({\bf 3}, {\bf 1}, X_{D_{a}} \right),\quad   a=1,2 
\end{eqnarray}
and the third family is assigned to  $SU(3)_L$ triplet:
\begin{eqnarray}
Q_{3 L} &=&  \left(t,b,T\right)_{L}^T
\ \sim \left({\bf 3}, {\bf 3}, X_{Q_{3}} \right),\quad t_{R}\sim \left({\bf 3}, {\bf 1}, X_{t} \right),\nonumber \\
b_{R}&\sim& \left({\bf 3}, {\bf 1}, X_{b} \right), \quad T_{R}\sim \left({\bf 3}, {\bf 1}, X_{T} \right) 
\end{eqnarray}
Furthermore, the Yukawa Lagrangian for quarks:
\begin{eqnarray}\label{Yukawaquarks}
-\mathcal{L}_{Y}^{\text{quarks}} &=& f_{ab}^{u}\: \overline{Q_{aL}}\; \rho^{*}\: u_{bR}
+  f^{d}_{ab}\: \overline{Q_{aL}}\; \eta^{*}\: d_{bR}  \nonumber \\
&+& f^{D}_{ab}\: \overline{Q_{aL}}\;  \chi^{*}\:  D_{bR}
+ f^{b}\:  \overline{Q_{3L}}\; \rho\:  b_{R}\nonumber \\
&+& f^{t}\: \overline{Q_{3L}}\;  \eta\:  t_{R}
+ f^{T}\:  \overline{Q_{3L}}\; \chi\:  T_{R} 
+ \text{h.c.}
\end{eqnarray}
The Yukawa Lagrangian for leptons:
\begin{eqnarray}\label{Yukawaleptons}
-\mathcal{L}_{Y}^{\text{leptons}} &=& F_{ij}^{e}\: \overline{\psi_{iL}}\; \rho\: e_{jR}
+  F^{E}_{ij}\: \overline{\psi_{iL}}\;\chi \: E_{jR} + \text{h.c.} 
\end{eqnarray}
On the other hand, for  $\beta=0$ we have leptons and the third quark family  in anti-triplets,
while the first two quark families in triplets. 
In this case, the leptonic sector of the model is \cite{simplest}
\begin{eqnarray}
\psi_{i L} &=&  \left(e^{-}_i,-\nu_i,E_i\right)_{L}^T
\ \sim \left( {\bf 1},{\bf 3^*}, X_{\ell_{i}} \right),\nonumber \\  
e_{i_{R}}^{-} &\sim&   \left( {\bf 1},{\bf 1}, X_{e_{i}} \right)   ,\quad  E_{i_{R}} \sim   \left( {\bf 1},{\bf 1}, X_{E_{i}} \right)
\end{eqnarray}
where $i=1,2,3$. \\
The two first quark families form $SU(3)_L$ triplets
\begin{eqnarray}
Q_{a L} &=&  \left(u_a,d_a,U_{a}\right)_{L}^T
\ \sim \left({\bf 3}, {\bf 3}, X_{Q_{a}} \right), \nonumber \\
u_{a_{R}} &\sim& \left({\bf 3}, {\bf 1}, X_{u_{a}} \right), \quad  
d_{a_{R}} \sim \left({\bf 3}, {\bf 1}, X_{d_{a}} \right),\nonumber \\
U_{a_{R}} &\sim& \left({\bf 3}, {\bf 1}, X_{U_{a}} \right),\quad   a=1,2
\end{eqnarray}
and the third family is assigned to  $SU(3)_L$ anti-triplet:
\begin{eqnarray}
Q_{3 L} &=&  \left(b,-t, T\right)_{L}^T
\ \sim \left({\bf 3}, {\bf 3^* }, X_{Q_{3}} \right),\quad t_{R}\sim \left({\bf 3}, {\bf 1}, X_{t} \right),\nonumber \\
b_{R}&\sim& \left({\bf 3}, {\bf 1}, X_{b} \right), \quad T_{R}\sim \left({\bf 3}, {\bf 1}, X_{T} \right) 
\end{eqnarray}
Finally, for this version  the Yukawa Lagrangian for quarks is:
\begin{eqnarray}\label{Yukawaquarks}
-\mathcal{L}_{Y}^{\text{quarks}} &=& f_{ia}^{u}\: \overline{Q_{iL}}\; \eta\: u_{aR}
+  f^{d}_{ia}\: \overline{Q_{iL}}\; \rho\: d_{aR}  \nonumber \\
&+& f^{U}_{ia}\: \overline{Q_{iL}}\;  \chi\:  U_{aR}
+ f^{a}\:  \overline{Q_{3L}}\; \eta^*\:  d_{aR}\nonumber \\
&+& f^{t}_{a}\: \overline{Q_{3L}}\;  \rho^*\: u_{aR}
+ f^{U}_{a}\:  \overline{Q_{3L}}\; \chi^*\: U_{aR} 
+ \text{h.c.}
\end{eqnarray}
with $i= 1,2; a = 1,2,3$; $u_{aR} = u_R, c_R, t_R$; $d_{aR} = d_R, s_R, b_R$ and $U_{aR} = U_{1R}, U_{2R}, T_R$. 
The Yukawa Lagrangian for leptons take 
the form:
\begin{eqnarray}\label{Yukawaleptons}
-\mathcal{L}_{Y}^{\text{leptons}} &=& F_{ij}^{e}\: \overline{\psi_{iL}}\; \rho\: e_{jR}
+  F^{E}_{ij}\: \overline{\psi_{iL}}\;\chi \: E_{jR} + \text{h.c.} 
\end{eqnarray}
Now, in order to obtain the different electric charges of the particles in these models, we will use classical and quantum constraints.
\section{Constraints from families replicas}
Since the SM particles and the exotic particles in the models present replicas between families, this allows us to reduce the number of hypercharge variables. For $\beta=1/\left(3\sqrt{3}\right)$:
\begin{eqnarray}
X_{Q_{1}} &=& X_{Q_{2}} \equiv X_{Q} \\
X_{\ell_{1}} &=& X_{\ell_{2}} = X_{\ell_{3}}  \equiv X_{\ell}\\
X_{u_{1}} &=& X_{u_{2}} = X_{t} \equiv X_{u}\\
\label{eqfor_d}X_{d_{1}} &=& X_{d_{2}} = X_{b} \equiv X_{d}\\
X_{e_{1}} &=& X_{e_{2}} = X_{e_{3}}  \equiv X_{e}\\
X_{E_{1}} &=& X_{E_{2}} = X_{E_{3}}  \equiv X_{E}\\  
\label{lastEQ}X_{D_{1}} &=& X_{D_{2}}  \equiv X_{D} 
\end{eqnarray}
and for $\beta= 0$, we obtain the same conditions, but instead of (\ref{lastEQ}), we have:
\begin{eqnarray} 
X_{U_{1}} &=& X_{U_{2}} =  X_{T} \equiv X_{U}  
\end{eqnarray}

\section{Constraints from $U(1)_{X}$ invariance}
From the $U(1)_{X}$ invariance of the Yukawa lagrangian we obtain for $\beta=1/\left(3\sqrt{3}\right)$:
\begin{eqnarray}
\label{eqfor_Q}X_{Q}&=&X_{d}-X_{\eta}\\
X_{Q}&=&X_{D}-X_{\chi}\\
X_{Q}&=&X_{u}-X_{\rho}\\
X_{Q_{3}}&=&X_{\eta}+X_{t}\\
X_{Q_{3}}&=&X_{\chi}+X_{T}\\
\label{eqfor_Q3}X_{Q_{3}}&=&X_{\rho}+X_{b}\\
\label{eqfor_e1} X_{e}&=& X_{\ell}-X_{\rho}\\
\label{eqfor_E1} X_{E}&=& X_{\ell}-X_{\chi}
\end{eqnarray}
and for $\beta=0$
\begin{eqnarray}
\label{eqfor_Q2}X_{Q}&=&X_{u}+X_{\eta}\\
\label{EqXQ}X_{Q}&=&X_{d} + X_{\rho}\\
X_{Q}&=&X_{\chi}+X_{U}\\
\label{eqXQ3}X_{Q_{3}}&=&X_{d}-X_{\eta}\\
X_{Q_{3}}&=&X_{u}-X_{\rho}\\
\label{eqfor_Q3_2}X_{Q_{3}}&=&X_{U}-X_{\chi}
\end{eqnarray}
the equations (\ref{eqfor_e1}) and (\ref{eqfor_E1}) are still being fulfilled for this case.

\section{Constraints from anomalies cancellation}
As it is known, an anomaly is a symmetry which has been conserved in the classical theory but is broken 
at the quantum level. In the context of quantum field theory involving chiral fermions, it is important to cancel  gauge anomalies in order to
obtain a renormalizable theory. In the present paper, we are focusing on the 
cancellation of gauge 
and mixed gauge-gravitational anomalies.
Thus, the  quantum restrictions arising from anomalies cancellation imply
{\footnotesize 
	\begin{eqnarray}
	\label{eq:Chiral-anomalies}
	&\left[\mathrm{\mathrm{SU}(3)}_{C} \right]^{2} \mathrm{\mathrm{U}(1)}_{X} \rightarrow & A_{C} = 3 \sum_{q}X_{q_{L}} - \sum_{q}X_{q_{R}}=0	\nonumber	\\
	&\left[\mathrm{\mathrm{SU}(3)}_{L} \right]^{2} \mathrm{\mathrm{U}(1)}_{X} \rightarrow & A_{L}  =  3\sum_{q}X_{q_{L}} +\sum_{\ell}X_{\ell_{L}} =0	\nonumber	\\
	&\left[\mathrm{Grav} \right]^{2}   \mathrm{\mathrm{U}(1)}_{X} \rightarrow & A_{\mathrm{G}}=
	3 \sum_{\ell, q}\left[X_{\ell_{L}}+3X_{q_{L}} \right]
	- \sum_{\ell, q}\left[X_{\ell_{R}}+3X_{q_{R}} \right]=0 \nonumber \\
	&\left[\mathrm{\mathrm{U}(1)}_{X} \right]^{3} \rightarrow & A_{X}=
	3 \sum_{\ell, q}\left[X_{\ell_{L}}^{3}+3X_{q_{L}}^{3} \right]	
	- \sum_{\ell, q}\left[X_{\ell_{R}}^{3}+3X_{q_{R}}^{3} \right] 	=0 \nonumber
	\end{eqnarray}
}
where $q_{L}$ and $\ell_{L}$ are the doublets, and $q_{R}$ and $\ell_{R}$ are the singlets, for the SM and exotic fields, then:
\begin{eqnarray}\label{eq:Chiral-anomalies}
A_{C} &=& 3\left\lbrace 2X_{Q} + X_{Q_{3}} \right\rbrace \nonumber \\&-& \left( 2 X_{u}+2X_{d}+2X_{D,Ux}+X_{t}+X_{b}+X_{T} \right)=0\nonumber  		\\ \\
\label{anomalieslep} A_{L} &=& 3\left\lbrace 2X_{Q} + X_{Q_{3}} \right\rbrace + 3 X_{\ell}=0 \\ \nonumber\ \\
\label{anomaliesgrav}A_{G} &=& 3\left( 3X_{\ell}\right)-\left\lbrace 3X_{e}+3X_{E} \right\rbrace =0
\end{eqnarray}
From equation (\ref{anomaliesgrav}), we have:
\begin{eqnarray}\label{AG}
A_{G} = 9 X_{\ell} - 3  X_{e} - 3 X_{E} = 0
\end{eqnarray}
\\
For $\beta=1/\left(3\sqrt{3}\right)$:\\
Using equations (\ref{Higgsvalues1}), (\ref{eqfor_e1}) and (\ref{eqfor_E1}) in (\ref{AG}), we obtain:
\begin{eqnarray}\label{lephiggs1}
X_{\ell} = -5 X_{\chi}   
\end{eqnarray}
\\
For $\beta=0$:\\
Replacing equations (\ref{Higgsvalues2}), (\ref{eqfor_e1}) and (\ref{eqfor_E1}) in (\ref{AG}), we have
\begin{eqnarray}\label{lephiggs2}
X_{\ell} = X_{\eta}   
\end{eqnarray}

\section{Results}

\subsection{For  $\beta=1/\left(3\sqrt{3}\right)$}
Substracting equations (\ref{eqfor_Q}) and (\ref{eqfor_Q3}) and using
(\ref{eqfor_d}) and (\ref{Higgsvalues}):
\begin{eqnarray}\label{QQ3}
X_{Q_{3}} =  X_{Q}-X_{\chi}
\end{eqnarray}
Replacing the equations (\ref{lephiggs1}) and (\ref{QQ3}) in (\ref{anomalieslep}), we obtain:
\begin{eqnarray}
X_{Q} =  2 X_{\chi}  
\end{eqnarray}
Then
\begin{eqnarray}
X_{Q_{3}} =  X_{\chi},\quad X_{D}=3X_{\chi}\nonumber \\
X_{T} =  0,\quad X_{d} =  -3X_{\chi},\quad X_{u} =  6X_{\chi} \nonumber \\
X_{e}=-9X_{\chi},\quad X_{E}=-6X_{\chi}
\end{eqnarray}
In this case the charge operators equation (\ref{charge}) is given by:
\begin{eqnarray}
\mathcal{Q}&=&\text{diag} \left( \pm 5/9, \mp 4/9, \mp 1/9 \right) + \dfrac{X}{9 X_{\chi}}\text{I}_{3\times 3} 
\end{eqnarray}

As it was previously explained, the upper sign corresponds to the fundamental representation (triplet) of 
the Gell-Mann matrices and the lower sign to the conjugate representation (anti-triplet).\\
For the lepton triplet, we obtain:
\begin{eqnarray}
\mathcal{Q} \psi_{L} &=&\left( \text{diag} \left(  5/9, - 4/9, - 1/9 \right) - \dfrac{5}{9}\text{I}_{3\times 3}\right) \psi_{L}\nonumber \\
\mathcal{Q} \psi_{L} &=&\text{diag} \left(  0, - 1, - 2/3 \right)  \psi_{L}
\end{eqnarray}
Thus, we find the quantization of the electric charge with the correct electric charges for leptons 
\begin{eqnarray}
\mathcal{Q}_{\nu_{e,\mu,\tau}} = 0,\quad \mathcal{Q}_{e,\mu,\tau} = -1,\quad \mathcal{Q}_{E,M,\mathcal{T}} = -2/3
\end{eqnarray}
and for the quark anti-triplet: 
\begin{eqnarray}
\mathcal{Q} Q&=&\text{diag} \left(  -1/3,  2/3, 1/3 \right)  Q
\end{eqnarray}
\begin{eqnarray}
\mathcal{Q}_{d,s} = -1/3,\quad \mathcal{Q}_{u,c} = 2/3,\quad \mathcal{Q}_{D_{1},D_{2}} = 1/3
\end{eqnarray}
for the quark triplet
\begin{eqnarray}
\mathcal{Q} Q_{3}&=&\text{diag} \left(  2/3,  -1/3, 0 \right)  Q_{3}
\end{eqnarray}
\begin{eqnarray}
\mathcal{Q}_{b} = -1/3,\quad \mathcal{Q}_{t} = 2/3,\quad \mathcal{Q}_{T} = 0
\end{eqnarray}

\subsection{For $\beta=0$}
Substracting equations (\ref{EqXQ}) and (\ref{eqXQ3}) and using (\ref{Higgsvalues2}):
\begin{eqnarray}\label{QQ3beta0}
X_{Q_{3}} =  X_{Q}
\end{eqnarray}
Replacing the equations (\ref{lephiggs2}) and (\ref{QQ3beta0}) in (\ref{anomalieslep}), we obtain:
\begin{eqnarray}
X_{Q} =  -\dfrac{1}{3} X_{\eta}  
\end{eqnarray}
Then
\begin{eqnarray}
X_{Q_{3}} = -1/3 X_{\eta},\quad X_{U,T}= -1/3 X_{\eta}\nonumber \\
\quad X_{d} =  2/3 X_{\eta},\quad X_{u} = -4/3 X_{\eta} \nonumber \\
X_{e}= 2 X_{\eta},\quad X_{E}= X_{\eta}
\end{eqnarray}
and 
\begin{eqnarray}
\mathcal{Q}&=&\text{diag} \left( \mp 1/2, \mp 1/2, 0 \right) - \dfrac{X}{2 X_{\eta}}\text{I}_{3\times 3} 
\end{eqnarray}
For the lepton triplet
\begin{eqnarray}
\mathcal{Q} \psi_{L} &=&\left( \text{diag} \left( -1/2, 1/2, 0 \right) - \dfrac{1}{2}\text{I}_{3\times 3}\right) \psi_{L}\nonumber \\
\mathcal{Q} \psi_{L} &=&\text{diag} \left(  -1, 0, - 1/2 \right)  \psi_{L}
\end{eqnarray}
and 
\begin{eqnarray}
\mathcal{Q}_{\nu_{e,\mu,\tau}} = 0,\quad \mathcal{Q}_{e,\mu,\tau} = -1,\quad \mathcal{Q}_{E,M,\mathcal{T}} = -1/2
\end{eqnarray}
For the quarks triplet
\begin{eqnarray}
\mathcal{Q} Q_{3}&=&\text{diag} \left(  -1/3,  2/3, -1/6 \right)  Q
\end{eqnarray}
\begin{eqnarray}
\mathcal{Q}_{b} = -1/3,\quad \mathcal{Q}_{t} = 2/3,\quad \mathcal{Q}_{T} = -1/6
\end{eqnarray}
Finally, for the quarks, anti- triplet :
\begin{eqnarray}
\mathcal{Q} Q&=&\text{diag} \left(  2/3,  -1/3, -1/6 \right)  Q_{3}
\end{eqnarray}
\begin{eqnarray}
\mathcal{Q}_{u,c} = 2/3,\quad \mathcal{Q}_{d,s} = -1/3,\quad \mathcal{Q}_{U_{1},U_{2}} = -1/6
\end{eqnarray}





\section{Conclusion}
In this work, we have considered two versions of the 331 model, with the particular feature of containing
extra leptons with fractional electric charges and non-standard electric charges for the new quarks. By considering constraints
from the classical and quantum level, we have shown, for both versions  $\beta=1/\left( 3\sqrt{3}\right)$
and $\beta=0$,  that the quantization of the electric charge can be obtained by using the Yukawa sector and the chiral anomalies
cancellation when the three families are taken together and independent of the neutrino, as happens in the others 331 versions. 
As it can be observed from our procedure, different $\beta$ values produce different constraint equations as a result of imposing
the $U(1)_{\chi}$ invariance of the Yukawa Lagrangian and the anomalies cancellation. This is due to
the parameter $\beta$ fixes the fermion representations in the multiplets of the group. In that sense, in our approach, we think that  the extension of the charge quantization for an arbitrary $\beta$ is not straightforward.\\
An analysis using, the general form of the electromagnetic currents under parity invariance for arbitrary beta, and the cancellation
of chiral anomalies for two specific values of the mentioned parameter, allows to obtain the quantization of the electric charge as
was shown in the reference \cite{Long}.


%
%



\end{document}